\documentclass[12pt,floatfix,showkeys,superscriptaddress,aps,prd,preprint]{revtex4}
\usepackage[latin1]{inputenc}
\usepackage[T1]{fontenc}
\usepackage{lmodern}
\usepackage{amsmath}
\usepackage{amssymb}
\usepackage{graphicx}
\usepackage{esint}
\usepackage{longtable}
\usepackage[english]{babel}
\usepackage{csquotes}
\usepackage{color}
\usepackage{slashed}
\usepackage{amsmath,latexsym}

\usepackage{slashed}

\usepackage{hyperref}
\hypersetup{colorlinks,breaklinks,
			citecolor=[rgb]{0,0.0,1.0},
            urlcolor=[rgb]{0.0,0.0,1.0},
            linkcolor=[rgb]{0,0.5,0.9}}

\def\be{\begin{equation}}
\def\ee{\end{equation}}
\def\bea{\begin{eqnarray}}
\def\eea{\end{eqnarray}}

\begin{document}

\title{Regular black hole solutions in $(2 + 1)$-dimensional $f(R,T)$ gravity coupled to nonlinear electrodynamics}

\author{Miguel A. S. Pinto}
\email{mapinto@fc.ul.pt}
\affiliation{Instituto de Astrof\'{i}sica e Ci\^{e}ncias do Espa\c{c}o, Faculdade de Ci\^{e}ncias da Universidade de Lisboa, Edif\'{i}cio C8, Campo Grande, P-1749-016 Lisbon, Portugal.}
\affiliation{Departamento de F\'{i}sica, Faculdade de Ci\^{e}ncias da Universidade de Lisboa,  Edif\'{i}cio C8, Campo Grande, P-1749-016 Lisbon, Portugal.}
\author{Roberto V. Maluf}
\email{r.v.maluf@fisica.ufc.br}
\affiliation{Universidade Federal do Cear\'a (UFC), Departamento de F\'isica,\\ Campus do Pici, Fortaleza - CE, C.P. 6030, 60455-760 - Brazil.}
\author{Gonzalo J. Olmo}
\email{gonzalo.olmo@uv.es}
\affiliation{Universidade Federal do Cear\'a (UFC), Departamento de F\'isica,\\ Campus do Pici, Fortaleza - CE, C.P. 6030, 60455-760 - Brazil.}
\affiliation{Departamento de F\'{i}sica Te\'{o}rica and IFIC, Centro Mixto Universidad de Valencia - CSIC. Universidad
de Valencia, Burjassot-46100, Valencia, Spain.}


\date{\today}

\begin{abstract}
In this paper, we investigate regular black hole solutions in the (2+1)-dimensional versions of General Relativity and $f(R, T)$  gravity, both coupled to nonlinear electrodynamics. By admitting that the matter content that generates such geometries satisfies the Maxwell limit condition, we obtain a class of regular black holes that give rise to new solutions and successfully reproduce particular cases found in earlier studies of (2+1)-dimensional General Relativity. Moreover, we discover the first regular black hole solutions in (2+1)-dimensional $f(R, T)$ gravity and explore both qualitatively and quantitatively the non-conservation of the energy-momentum tensor present in those solutions. 
\end{abstract}

\keywords{Regular black holes, $f(R,T)$ gravity, Nonlinear electrodynamics, $(2+1)$-dimensional spacetime}

\maketitle


\section{Introduction}
Since April 2019, after the Event Horizon Telescope (EHT) collaboration announced the first direct image of a compact object at the center of Messier 87, we have very strong evidence that black holes (BHs), exotic astrophysical objects theoretically predicted by Einstein's General Relativity (GR), do exist \cite{ETH_2019_1,ETH_2019_2,ETH_2019_3,ETH_2019_4,ETH_2019_5,ETH_2019_6}. Furthermore, at the beginning of 2022, this indication received even further reinforcement when the same collaboration announced a second image, this time capturing Sagittarius A$^*$, the compact object at the center of our galaxy \cite{ETH_2022_1,ETH_2022_2,ETH_2022_3,ETH_2022_4,ETH_2022_5,ETH_2022_6}. Since then, new static and animated images, including maps of the light polarization patterns, have been published \cite{EventHorizonTelescope:2021bee,EventHorizonTelescope:2021srq}, offering fundamental new information about physical processes going on in these extreme environments. 

Though BHs appear as the most popular candidates to interpret these observations, not only because of their historical importance but also due to their consistency with observational data, it is undeniable that they involve conceptual issues that cannot be easily ignored. The first one is the presence of horizons, which ultimately leads to the famous BH information loss paradox and an apparent conflict with the foundations of quantum theory. The second is the existence of (physical) singularities, which may imply pathologies in the space-time structure (curvature divergences) and spoil the predictability of the laws of Physics \cite{Penrose:1964wq,Hawking:1966vg,Hawking:1970zqf,Hawking:1976ra}. In addition, some metrics that are assumed to describe these astrophysical objects also admit closed time-like curves, which are inconsistent with our notion of causality \cite{Calvani:1978xj,deFelice:1979jnj,deFelice:1980xm}. In fact, these last two peculiar aspects of BH geometries are, among others, one of the main drivers for the pursuit of a consistent quantum theory of gravity, which is expected to cure such pathologies by somehow ``smoothing out'' the solutions \cite{Bambi:2013ufa}.

Alternative compact objects offer a plausible way to circumvent the above issues. Some of the candidates that have received substantial attention in the last decades are boson stars \cite{Kaup:1968zz}, gravastars \cite{Mazur:2001fv}, and Proca stars \cite{Brito:2015pxa}, all of which represent horizonless compact objects. Nonetheless, regular BHs (RBHs) (see \cite{Lan:2023cvz} for a review) have also received much attention. Indeed, in the absence of a fully consistent quantum theory of gravity, RBHs constitute a unique window to explore the phenomenology one would expect in gravitational scenarios that lead to robust outcomes under gravitational collapse.

In general terms, RBHs are  not vacuum solutions of some modified gravitational field equations. Instead, they are typically obtained as solutions of Einstein's theory (or any other gravity theory) coupled to some exotic matter source. The first regular solution in the literature is attributed to James Bardeen \cite{Bardeen} and was constructed as an {\it ad hoc} modification of the  Schwarzschild solution, adding a new scale in the $r\to 0$ limit. Yet, this solution and others inspired by it \cite{Borde:1994ai, Barrabes:1995nk, Mars:1996khm} lacked physical motivation because of their {\it ad hoc} character. We had to wait until the early 2000s to make some progress in this direction, when Ay\'{o}n-Beato and Garcia solved this issue by successfully interpreting Bardeen's BH as a solution of GR minimally coupled to a magnetic monopole in the context of nonlinear electrodynamics (NED) \cite{Ayon-Beato:2000mjt}. This interpretation established a common approach to construct (classical) RBHs, that is, to consider GR as the theory of gravity and the matter content as described by NED \cite{Bronnikov:2000vy,Burinskii:2002pz,Balart:2014cga,Ma:2015gpa,Toshmatov:2017zpr,Rodrigues:2018bdc}. In practical terms, these RBH solutions are generated by vacuum polarization effects due to the presence of a nonlinear electromagnetic field, which, in turn, may be engendered by effective quantum effects resulting from a more fundamental description of gravity.

Likewise, RBH solutions have also been discovered in (2+1)-dimensional GR coupled to NED \cite{Cataldo:2000ns, He:2017ujy}. In fact, the first and most famous regular solution in this lower-dimensional space-time, the so-called BTZ solution \cite{Banados:1992wn, Banados:1992gq}, was obtained by introducing a negative cosmological constant $\Lambda$ into the Einstein field equations. The fact that  the (2+1) dimensional scenario serves as a theoretical laboratory to test subtle aspects of classical and quantum gravity \cite{Carlip:1995zj, Padmanabhan:2010zzb, Carlip:2023nwa} is a key motivation to explore RBH solutions generated by modified gravitational dynamics in interaction with NEDs \cite{Bueno:2021krl,
Bueno:2022ewf,
Bueno:2025jgc}. The existence of such solutions beyond the domain of classical GR reinforces the robustness of these entities as plausible compact objects despite their nontrivial causal structure.  

Among the many alternative gravity theories being currently explored in the literature, we find $f(R, T)$ theories as particularly relevant for our discussion \cite{Harko:2011kv}, as they are able to capture relevant phenomenology expected in quantum gravitational scenarios \cite{Harko:2018ayt,Dzhunushaliev:2013nea, Yang:2015jla}. In particular, theories of this type, with non-minimal matter-curvature couplings, are expected to induce energy exchanges between the gravitational and matter sectors, possibly leading to violations in the conservation of the energy-momentum tensor of the matter. The impact that this new physics could have on the existence of event horizons and regular compact objects is thus one of the reasons that motivates this choice of theories. However, $f(R,T)$ possesses a rather complicated mathematical structure due to the presence of the trace of the energy-momentum tensor $T$. In fact, obtaining solutions in (3+1) dimensions, even for relatively simple scenarios, is usually very demanding if not impossible. Therefore, in this paper, we reduce the dimensionality of the problem to investigate the consequences of simply adding a linear $T$ term to the (2+1)-dimensional Einstein-Hilbert action with a negative cosmological constant $\Lambda$, aiming to obtain analytical solutions. Indeed, this approach retains the freedom of the modified gravity sector within reasonable bounds while avoiding excessive technical complications. As we will see, our approach yields new RBH solutions and recovers results already known in the literature in the appropriate limits.

The paper is organized as follows. In Sec. \ref{review}, we derive the modified field equations of $f(R,T)$ gravity in (2+1) dimensions, coupled to nonlinear electrodynamics, along with its gauge equations. In Sec. \ref{Results}, we propose a generalized electric field that satisfies the Maxwell asymptotic limit and then derive the corresponding forms for the nonlinear electrodynamics Lagrangian, as well as the RBH solutions for both GR and $f(R,T)$ cases. The conclusions are presented in Sec. \ref{conclusion}.

\section{Equations of motion in (2+1)-dimensional $f(R,T)$ gravity \label{review}}
In the present section, we begin by introducing the general equations of $f(R,T)$ gravity in (2+1)-dimensions coupled to NED. Then, we particularize the equations by taking a concrete $f(R, T)$ function, and specifying the metric \textit{ansatz} and the form of the Faraday-Maxwell tensor $F_{\mu\nu}$. Throughout this work, we adopt the natural units system, i.e., $\hbar = c = 1$.

\subsection{General equations}
The action for $f(R,T)$ gravity in 2+1 dimensions minimally coupled to NED can be formulated as 

\begin{equation}\label{eq:fRTaction-original}
    S=\int d^{3}x\sqrt{-g}\left[\frac{1}{2\kappa^{2}}f(R,T)+L(F)\right],
\end{equation}
where $\kappa^2=8\pi \Tilde{G}$, with $\Tilde{G}$ being the Newtonian gravitational constant in two spatial dimensions, $g$ is the determinant of the metric tensor $g_{\mu \nu}$, and $f(R,T)$ is a mathematically well-behaved function of the Ricci scalar, $R$, and of the trace of the energy-momentum tensor, $T=g^{\mu \nu}T_{\mu \nu}$ 
The latter is defined as
\begin{equation}
T_{\mu\nu}\equiv-\frac{2}{\sqrt{-g}}\frac{\delta\left[\sqrt{-g}L(F)\right]}{\delta g^{\mu\nu}},
\end{equation}
where $L(F)$ is the NED Lagrangian density, with $F\equiv F_{\mu \nu}F^{\mu\nu}$ \cite{Cataldo:2000ns,He:2017ujy}. The corresponding electromagnetic strength tensor, written in terms of the potential $A_{\mu}$, is given by
\begin{equation}
\label{electro_field}
    F_{\mu \nu} = \partial_{\mu}A_{\nu}-\partial_{\nu}A_{\mu},
\end{equation}
in terms of which the energy-momentum tensor becomes
\begin{equation}
\label{energy-momentum2}
    T_{\mu \nu} = L(F)g_{\mu\nu}-4L_{F}F_{\mu}^{\ \alpha}F_{\nu\alpha} \,
\end{equation}
where $L_{F}$ represents the derivative of $L(F)$ with respect to $F$.

On the other hand, the variation of Eq. (\ref{eq:fRTaction-original}) with respect to the metric yields the modified field equations of $f(R,T)$ gravity
\begin{equation}
f_{R}R_{\mu\nu}-\frac{1}{2}g_{\mu\nu}f(R,T)+\left(g_{\mu\nu}\Box-\nabla_{\mu}\nabla_{\nu}\right)f_{R}=\kappa^{2}T_{\mu\nu}-f_{T}\left(T_{\mu\nu}+\Theta_{\mu\nu}\right),\label{EoMg}
\end{equation}
where $\nabla_\mu$ is the covariant derivative, $\square\equiv\nabla^\alpha\nabla_\alpha$ is the D`Alembertian, with $f_R$ and $f_T$ denoting the partial derivatives of the $f(R,T)$ function with respect to $R$ and $T$, respectively. Moreover, the auxiliary tensor $\Theta_{\mu \nu}$ has the following form
\begin{equation}\label{eq:Theta-varT}
    \Theta_{\mu\nu}\equiv g^{\rho\sigma}\frac{\delta T_{\rho\sigma}}{\delta g^{\mu\nu}}=-L(F)g_{\mu\nu}+2\left(L_{F}-4L_{FF}F\right)F_{\mu}^{\ \alpha}F_{\nu\alpha},
\end{equation}
where $L_{FF}$ represents the second-order derivative of $L(F)$ with respect to $F$. Since our spacetime is three-dimensional, for the usual electrodynamics (with $L(F)=-F$) we obtain $\Theta_{\mu\nu}=-T_{\mu\nu}+2 F_{\mu}^{\ \alpha}F_{\nu\alpha}$ \cite{Harko:2011kv}. Additionally, taking the trace of Eq. (\ref{EoMg}) yields
\begin{equation}
    f_{R}R-\frac{3}{2}f+2\Box f_{R}=\left(\kappa^{2}-f_{T}\right)T-f_{T}\Theta,\label{EoMtrace}
\end{equation}
where $\Theta\equiv g^{\mu\nu}\Theta_{\mu\nu}=-3L+2F(L_{F}-4L_{FF}F)$ and $T=3L-4L_{F}F$.

A consequence of the general covariance of $f(R,T)$ gravity is the existence of a Bianchi identity, which can be obtained by taking the covariant divergence of Eq. \eqref{EoMg}. The result can be expressed as 
\begin{equation} \label{eq:conserv-general}
    \nabla^{\mu} T_{\mu\nu} = \frac{1}{\kappa^2 - f_T} \left[\left(T_{\mu\nu} + \Theta_{\mu\nu}\right) \nabla^\mu f_T + f_T \nabla^{\mu}\left( \Theta_{\mu\nu} -\frac{1}{2}g_{\mu\nu} T\right) \right],
\end{equation}
where we have used $(\Box\nabla_{\nu}-\nabla_{\nu}\Box)f_{R}=R_{\mu\nu}\nabla^{\mu}f_{R}$, and $\nabla^\mu \left(R_{\mu\nu} - \frac{1}{2} g_{\mu\nu} R\right) = 0$. Equation (\ref{eq:conserv-general}) indicates that the conservation of the energy-momentum tensor is not guaranteed for generic $f(R,T)$ theories. Though this may raise some alarms at first, one should note that theories with matter-curvature couplings are expected to induce energy-momentum exchanges between the matter fields and geometry, possibly allowing for matter creation by intense gravitational fields and also involving non-linear interactions among the components of the matter sector. Effects of this type may be in conflict with violations of the equivalence principle \cite{Olmo:2006zu,Barrientos:2018cnx,Bertolami:2006js,Damour:1996xt,Damour:1994zq,Damour:2010rp} and can be seen as signals of new physics beyond the current standard model. 
 In fact, the non-conservation of $T_{\mu\nu}$ is naturally described in the framework of the irreversible thermodynamics of open systems, in which matter creation results from an irreversible flow of  energy from the gravitational sector \cite{Harko:2014pqa,Pinto:2022tlu,Cipriano:2023yhv} (see \cite{Pinto:2023phl} for a brief review on this topic). The impact that this new physics could have on the existence of event horizons and regular compact objects is thus one of the main reasons that motivate our analysis.
 
At last, varying Eq. (\ref{eq:fRTaction-original}) with respect to the potential $A_{\mu}$ yields a generalized version of the NED equations,
\begin{equation}
\nabla^{\mu}\left[L_{F}F_{\mu\nu}-\frac{f_{T}}{2\kappa^{2}}\left(L_{F}+4L_{FF}F\right)F_{\mu\nu}\right]=0 ,\label{EoMGauge}
\end{equation}
with an explicit new coupling to the gravitational sector via the function $f_T$ which induces a dependence on the second derivative of the NED Lagrangian, thus adding new effects in the matter sector. 

\subsection{Choice of the $f(R,T)$ function}
In order to remain as close as possible to standard GR, we consider an extension involving a linear dependence on $T$ in the gravity Lagrangian, which leads to 
\begin{equation}
\label{function_fRT}
    f(R,T)=R-2\Lambda+\lambda T,
\end{equation}
where $\Lambda$ is a cosmological constant term, which can also be written using the anti-de Sitter (AdS) radius $\ell$ as  $\Lambda= -1/\ell^{2}$. As a brief remark, we note that the cosmological constant $\Lambda$ is crucial for ensuring the existence of black hole solutions with horizons in (2+1) dimensions. As is well known from the BTZ solution \cite{Banados:1992wn,Banados:1992gq}, without $\Lambda$ the spacetime remains locally flat, and no black hole geometry can be supported. Thus, $\Lambda$ provides the necessary curvature for non-trivial solutions, yielding the asymptotically AdS background required for our analysis.

Furthermore, regarding the choice of the $f(R,T)$ function used in this work, it was recently argued in Ref. \cite{Akarsu:2023lre} that a function of this type, where $R$ and $T$ are not intertwined, leads to the presence of a non-minimal matter interaction instead of a genuine non-minimal matter-curvature coupling, despite the energy-momentum tensor not being conserved. Hence, such an additive term can be seen not as a true modification of gravity but as a modification of the matter sector itself. Regardless of the interpretation, here we focus on the phenomenology it produces.

By introducing Eq. \eqref{function_fRT} into  Eqs. \eqref{EoMg} and (\ref{EoMGauge}), the field equations boil down to 
\begin{equation}
    R_{\mu\nu}-\frac{1}{2}g_{\mu\nu}R+\Lambda g_{\mu\nu}=\left(\kappa^{2}-\lambda\right)T_{\mu\nu}+\frac{\lambda}{2}g_{\mu\nu}T-\lambda\Theta_{\mu\nu},\label{EoMmetric}
\end{equation}
and
\begin{equation}
\nabla^{\mu}\left[\left(1-\frac{\lambda}{2\kappa^{2}}\right)L_{F}F_{\mu\nu}-\frac{2\lambda}{\kappa^{2}}L_{FF}F F_{\mu\nu}\right]=0.\label{EoMGauge_particular}
\end{equation}
Moreover, the conservation equation in Eq. (\ref{eq:conserv-general}) simplifies to
\begin{equation}\label{Energy2}
   \nabla^{\mu}T_{\mu\nu}= \frac{\lambda}{\kappa^2-\lambda} \nabla^{\mu}\left(\Theta_{\mu\nu}-\frac{1}{2}g_{\mu\nu}T\right).
\end{equation}
It is worth noting that in the Maxwell case the energy conservation equation reduces to  
\begin{equation}\label{Energy3}
   \left(1-\frac{\lambda}{2\kappa^{2}}\right)\nabla^{\mu}T_{\mu\nu}= 0,
\end{equation}  
ensuring that the conservation of the energy-momentum tensor is preserved for $\lambda \neq 2\kappa^{2}$. However, the case $\lambda = 2\kappa^{2}$ may introduce inconsistencies in the field equations, as we will see in what follows.

Assuming a circularly symmetric space-time, our  \textit{Ansatz} for the metric line element is
\begin{equation}
\label{metric_ansatz}
ds^{2}=-b(r)dt^{2}+\frac{1}{b(r)}dr^{2}+r^{2}d\theta^{2},
\end{equation}
where $b(r)$ is an unknown function of the variable $r$. 
Besides, for simplicity, we restrict the Faraday-Maxwell tensor $F_{\mu\nu}$ such that the only nonzero component is the electric field (that depends exclusively on the radial coordinate) \cite{Cataldo:2000ns}, neglecting the effects of potential magnetic charges, which implies that 
\begin{equation}
\label{ansatz_F}
F_{\mu\nu}=E(r)\left(\delta_{\mu}^{t}\delta_{\nu}^{r}-\delta_{\mu}^{r}\delta_{\nu}^{t}\right).
\end{equation}
With Eq. \eqref{ansatz_F}, it is easy to see that the invariant $F$ is given by 
\begin{equation}
F=-2E^{2}(r).
\end{equation}

With these assumptions, we can find a first integral of Eq. (\ref{EoMGauge_particular}) in the form
\begin{equation}
\label{gauge_1}
rE(r)\left[\left(1-\frac{\lambda}{2\kappa^{2}}\right)L_{F}+\frac{4\lambda}{\kappa^{2}}E^{2}(r)L_{FF}\right]=e,
\end{equation}
where $e$ is an integration constant. In this regard, it is worth noting that the choice $\lambda=2\kappa^2$ in the Maxwell case may lead to inconsistencies in the field equations, which can be avoided if one redefines $e$ in that case as $e\equiv-q\left(1-\frac{\lambda}{2\kappa^{2}}\right)$. We will thus adopt this form for $e$, such that the electric field becomes $E(r)=q/r$ for arbitrary $\lambda$, with $q$ representing the standard electric charge.

Taking now the modified Einstein equations (\ref{EoMmetric}) with our metric \textit{ansatz}, Eq. \eqref{metric_ansatz}, we
get the following $(tt)$ and $(\theta\theta)$ components, respectively
\begin{align}
\frac{b'(r)}{2r}+\Lambda & =\kappa^{2}\left(L(F)+4E^{2}(r)L_{F}\right)\nonumber \\
 & +\lambda\left(\frac{3}{2}L(F)+2E^{2}(r)L_{F}+16E^{4}(r)L_{FF}\right),\label{G00}
\end{align}
\begin{equation}
\frac{b''(r)}{2}+\Lambda=\kappa^{2}L(F)+\lambda\left(\frac{3}{2}L(F)+4E^{2}(r)L_{F}\right),\label{Gphiphi}
\end{equation}
where the prime $(')$ stands for the derivative with respect to the radial coordinate $r$. Note that we did not include the $(rr)$ component of the field equations, as it coincides with the $(tt)$ component up to a multiplicative factor. In addition, one can prove that the $(\theta \theta)$ component is not an independent equation by computing the derivative of $(tt)$ component, Eq. (\ref{G00}), and then using the equation of motion for the gauge field, Eq. (\ref{EoMGauge}). Thus, we only need to integrate Eq. (\ref{G00}), which formally leads to
\begin{align}
b(r) & =-M-\Lambda r^{2}+2\kappa^{2}\int r\left[L\left(F\right)+4E^{2}(r)L_{F}\right]dr\nonumber\\
 & +\lambda\int r\left[3L\left(F\right)+4E^{2}(r)\left(L_{F}+8E^{2}(r)L_{FF}\right)\right]dr\label{bsol}.
\end{align}
In order to make further progress with the determination of the metric function $b(r)$, we need to find the radial dependence of $E(r)$, which depends on the NED Lagrangian according to (\ref{gauge_1}). To proceed, we apply the chain rule, 
\begin{equation}
    L_F = \frac{dL}{dF}=\frac{dL}{dr}\frac{dr}{dF}= \frac{L'}{F'} \ ,
\end{equation}
and substitute the result in Eq. \eqref{gauge_1}, finding that 
\begin{equation}
\frac{L'}{4E'}+\frac{\lambda}{4\kappa^{2}}\left(\frac{L'}{2E'}-\frac{EL''}{E'^{2}}+\frac{EL'E''}{E'^{3}}\right)=\frac{q}{r}\left(1-\frac{\lambda}{2\kappa^{2}}\right).\label{eq:L-Econstrain}
\end{equation}
Moreover, the conservation equation, although useful in a follow-up discussion, does not provide additional information as it is proportional to Eq. \eqref{EoMGauge_particular}.

Thus, we end up with only two independent equations, Eqs. (\ref{bsol}) and (\ref{eq:L-Econstrain}), and three unknown functions, $b(r)$, $L(r)$, and $E(r)$. Typically, $L(F(r))$ is initially provided, and then one can solve the two equations to obtain $b(r)$ and $E(r)$ \cite{Cataldo:2000ns}. Another possibility is to construct RBH solutions first by specifying the form of $b(r)$ and then derive the specific forms of the associated NED Lagrangian and electric field, as done in Ref. \cite{He:2017ujy}. However, here we will follow an alternative approach by providing a rather general parametrized form for the electric field, which will allow us to obtain analytical solutions for $b(r)$ as well as agreement with previous results in the literature. The details appear next.

\section{Regular black hole solutions\label{Results}}

In this section, we first propose a generalized electric field capable of satisfying the Maxwell asymptotic limit and then derive the associated forms of the Lagrangian for nonlinear electrodynamics and the RBH solutions, according to the field equations, for two cases: GR ($\lambda=0$) and $f(R,T)$ gravity ($\lambda\neq 0$).

\subsection{Choice of the electric field}

Looking at Eq. (\ref{eq:L-Econstrain}), we note that it is a nonlinear differential equation for the electric field $E(r)$. Still, it preserves a linear structure for the Lagrangian density $L(r)$, which, in principle, can simplify the task of discovering new black hole solutions. So, our proposal is to adjust the electric field based on the usual Maxwell asymptotic limit and subsequently determine $L(r)$ and $b(r)$.

The usual electric profile given by $1/r$ in $(2+1)$ dimensions can be generalized to arbitrary powers of $r$ while still preserving the desired asymptotic Maxwell behavior in many different ways. The physical motivation behind such a change is related to the idea that in the strong gravity regime, electrodynamics might undergo modifications that should decay sufficiently fast in weaker regimes. Since this reasoning is independent of the dimensionality of the spacetime, instead of considering a particular NED Lagrangian, here we aim to provide a specific alternative form for the electric field that alters electric interactions at a characteristic length scale $a$ while keeping the usual Maxwell behavior in the asymptotic limit $r \rightarrow \infty$. Accordingly, inspired by these considerations and by solution (29) of Ref. \cite{Cataldo:2000ns}, we propose the following functional form for the electric field:
\begin{equation}
E(r)=\frac{q r^{\alpha}}{\left(r^{\beta}+a^{\beta}\right)^{\frac{\alpha+1}{\beta}}},\label{electricfield}
\end{equation}where we consider two arbitrary real positive exponents (including zero), $\alpha$ and $\beta$, to keep it general. Additionally, the length dimensions of the parameters $a$ and $q$ are $[q] = -1/2$ and $[a] = 1$ in natural units. Clearly, when $r \rightarrow \infty$, the electric field tends to $q/r$.

The simplest case is when $\alpha=a=0$ and $\beta=1$, and it is characterized by the Maxwell electric field 
\begin{equation}
E(r)=\frac{q}{r}.\label{E-Maxwell}
\end{equation}
Inserting (\ref{E-Maxwell}) in Eq. (\ref{eq:L-Econstrain}), the Lagrangian density takes the
form
\begin{equation}
L(r)=\frac{2q^{2}}{r^{2}}-\frac{2c_{1}\lambda}{\left(2\kappa^{2}+3\lambda\right)}\frac{1}{r^{\frac{3}{2}+\frac{\kappa^{2}}{\lambda}}}+c_{2},
\end{equation}
where $c_{1}$ and $c_{2}$ are integration constants. Choosing $c_{1}=c_{2}=0$,
we recover the usual Maxwell Lagrangian
\begin{equation}
L(r)=-F=2E^{2}(r)=\frac{2q^{2}}{r^{2}}.
\end{equation}
Substituting the above results into Eq. (\ref{bsol}), we obtain the corresponding metric function
\begin{equation}
b(r)=-M+\frac{r^{2}}{\ell^2}-\left(2\kappa^{2}-\lambda\right)q^{2}\ln\left( \frac{r^2}{\ell^2}\right),
\end{equation}
which, for $\lambda=0$, represents the static charged BTZ solution in the context of GR \cite{Banados:1992wn}. Note that the $\lambda$-corrections have no perceptible physical effect, as one can absorb it through a redefinition of the electric charge. Nonetheless, if we set $\lambda = 2\kappa^2$, we recover the uncharged static BTZ black hole, with $b(r)= -M-\Lambda r^{2}$ \cite{Banados:1992gq}.

Let us now investigate with more generality the solutions generated by the electric field function (\ref{electricfield}) in both GR ($\lambda=0$) and $f(R,T)$ gravity ($\lambda\neq 0$).


\subsection{GR solutions ($\lambda=0$)}

Substituting the expression for the electric field, Eq. (\ref{electricfield}), into Eq. (\ref{eq:L-Econstrain}), with $\lambda=0$, results in a first-order linear differential equation for the function $L(r)$. Upon integrating the equation, we obtain the following expression for the Lagrangian of the nonlinear electrodynamics as a function of the radial coordinate
\begin{align}
   L(r)=&	-\frac{2q^{2}}{a^{2}}+\frac{4q^{2}r^{\alpha-1}}{(\alpha+\beta-1)\left(r^{\beta}+a^{\beta}\right)^{\frac{\alpha+1}{\beta}}}\left[\frac{\alpha(\alpha+\beta-1)}{\alpha-1}\,_{2}F_{1}\left(1,-\frac{2}{\beta};\frac{\alpha+\beta-1}{\beta};-\left(\frac{r}{a}\right)^{\beta}\right)\right.\nonumber\\
	&\left.-\left(\frac{r}{a}\right)^{\beta}\,_{2}F_{1}\left(1,\frac{\beta-2}{\beta};\frac{\alpha-1}{\beta}+2;-\left(\frac{r}{a}\right)^{\beta}\right)\right],\label{Lr}
\end{align}
where the function $\,_{2}F_{1}\left(\text{a},\text{ b; c; z}\right)$ is Gauss's hypergeometric function \cite{weisstein_hypergeometric}. The integration constant was chosen such that, for $\alpha=3$ and $\beta=2$, $E(r)$ and $L(r)$ coincide with the expressions of a particular case previously studied in Ref. \cite{Cataldo:2000ns}, that is
\begin{equation}
E(r)=\frac{q r^{3}}{\left(r^{2}+a^{2}\right)^{2}},
\end{equation}and 
\begin{equation}
L(r)=\frac{2q^{2}\left(r^{2}-a^{2}\right)}{\left(r^{2}+a^{2}\right)^{2}}.\label{L2000}
\end{equation}
Note that the expression (\ref{L2000}) for $L(r)$ is asymptotically Maxwell, as $L(r)\rightarrow 2q^2/r^2$ when $r$ is large. 

Next, by substituting Eqs. (\ref{electricfield}) and (\ref{Lr}) into Eq. (\ref{bsol}), with $\lambda=0$, and integrating the resulting expression, we determine the metric function 
\begin{align}
    b(r)	&=-M-\left(\Lambda+\frac{2\kappa^{2}q^{2}}{a^{2}}\right)r^{2}\nonumber\\
	&-4\kappa^{2}q^{2}\int^{r} d\bar{r}\frac{\bar{r}^{\alpha}}{\left(\bar{r}^{\beta}+a^{\beta}\right)^{\frac{\alpha+1}{\beta}}}\left[1-\left(\frac{\alpha+1}{\alpha-1}\right)\,_{2}F_{1}\left(1,-\frac{2}{\beta};\frac{\alpha+\beta-1}{\beta};-\left(\frac{\bar{r}}{a}\right)^{\beta}\right)\right].\label{bsolution1}
\end{align}
To guarantee the convergence of the hypergeometric function, we must have $\frac{\alpha+\beta-1}{\beta}>0$, which, in our case, implies $\alpha>1$ and $\beta>0$. 

It is worth pointing out that our generalization for regular $(2 + 1)$-dimensional black holes in Eq. (\ref{bsolution1}) actually represents an infinite family of solutions and encompasses several previously studied results in the literature as a subset of possible regular black hole geometries. For example,  it is straightforward to verify that for $\alpha=3$ and $\beta=2$, and by making the replacements $\kappa^{2}=8\pi$ and $q^{2}\rightarrow q^{2}/16\pi$, the expression (\ref{bsolution1}) reduces exactly to that obtained in \cite{Cataldo:2000ns} (see Eq. (28) in that reference), that is
\begin{equation}
b(r)=-M-\Lambda r^{2}-2\kappa^2 q^{2}\ln\left(\frac{r^2}{a^2}+1\right).
\end{equation}
Another example is when $\alpha=2$ and $\beta=1$. From Eqs. (\ref{electricfield}), (\ref{Lr}) and (\ref{bsolution1}), we find 
\begin{equation}
    E(r)=\frac{qr^{2}}{(r+a)^{3}},\ \ \ \  L(r)=\frac{2q^{2}(r-a)}{(r+a)^{3}},
\end{equation}and
\begin{equation}
    b(r)=-M-\Lambda r^{2}-4\kappa^{2}q^{2}\left[\ln\left(\frac{r}{a}+1\right)-\frac{r}{r+a}\right],
\end{equation}
which shares the structure of a model considered in Ref. \cite{He:2017ujy} that also represents one RBH solution obeying the asymptotic Maxwell behavior. Moreover, this RBH solution will reduce to a static charged BTZ black hole when $r\rightarrow\infty$ (with a shifted mass term).  

On the other hand, for $\alpha=\beta=2$, we obtain one new type of solution in the form
\begin{equation}
E(r)=\frac{qr^{2}}{(r^{2}+a^{2})^{3/2}},\ \ \ \ L(r)=\frac{2q^{2}\left[2r^{3}+4ra^{2}-(r^{2}+a^{2})^{3/2}\right]}{a^{2}(r^{2}+a^{2})^{3/2}},
\end{equation}and
\begin{equation}
b(r)=-M-\left(\Lambda+\frac{2\kappa^{2}q^{2}}{a^{2}}\right)r^{2}+4\kappa^{2}q^{2}\left[\frac{r}{a^{2}}\sqrt{r^{2}+a^{2}}+\ln\left(\frac{\sqrt{r^{2}+a^{2}}-r}{a}\right)\right]\label{bnew}.
\end{equation}
To the best of our knowledge, this model represents a new RBH solution. One can check that in the $r\gg a$ limit, the Lagrangian recovers Maxwell up to a constant, and the geometry tends to the static BTZ black hole with a modified cosmological constant. The regularity of the above solution can be verified by calculating the curvature invariants, namely, the Ricci $R$ and Kretschmann $K$ scalars for the function (\ref{bnew}), which leads to 
\begin{equation}
    R=6\Lambda+\frac{4\kappa^{2}q^{2}\left[3\left(r^{2}+a^{2}\right)^{3/2}-8a^{2}r-6r^{3}\right]}{a^{2}\left(r^{2}+a^{2}\right)^{3/2}},
\end{equation}
\begin{align}
    K	&=4\left\{ 3\Lambda^{2}+\frac{4\Lambda\kappa^{2}q^{2}\left[3\left(r^{2}+a^{2}\right)^{3/2}-8a^{2}r-6r^{3}\right]}{a^{2}\left(r^{2}+a^{2}\right)^{3/2}}\right.\\	&\left.+\frac{4\kappa^{4}q^{4}\left[41a^{2}r^{4}+33a^{4}r^{2}+3a^{6}+15r^{6}-4r\left(3r^{2}+4a^{2}\right)\left(r^{2}+a^{2}\right)^{3/2}\right]}{a^{4}\left(r^{2}+a^{2}\right)^{3}}\right\} .\nonumber
\end{align}
As is evident from these expressions, both Ricci and Kretschmann scalars are smooth everywhere.

\begin{figure}[!h]
\begin{center}
\begin{tabular}{ccc}
\includegraphics[height=7cm]{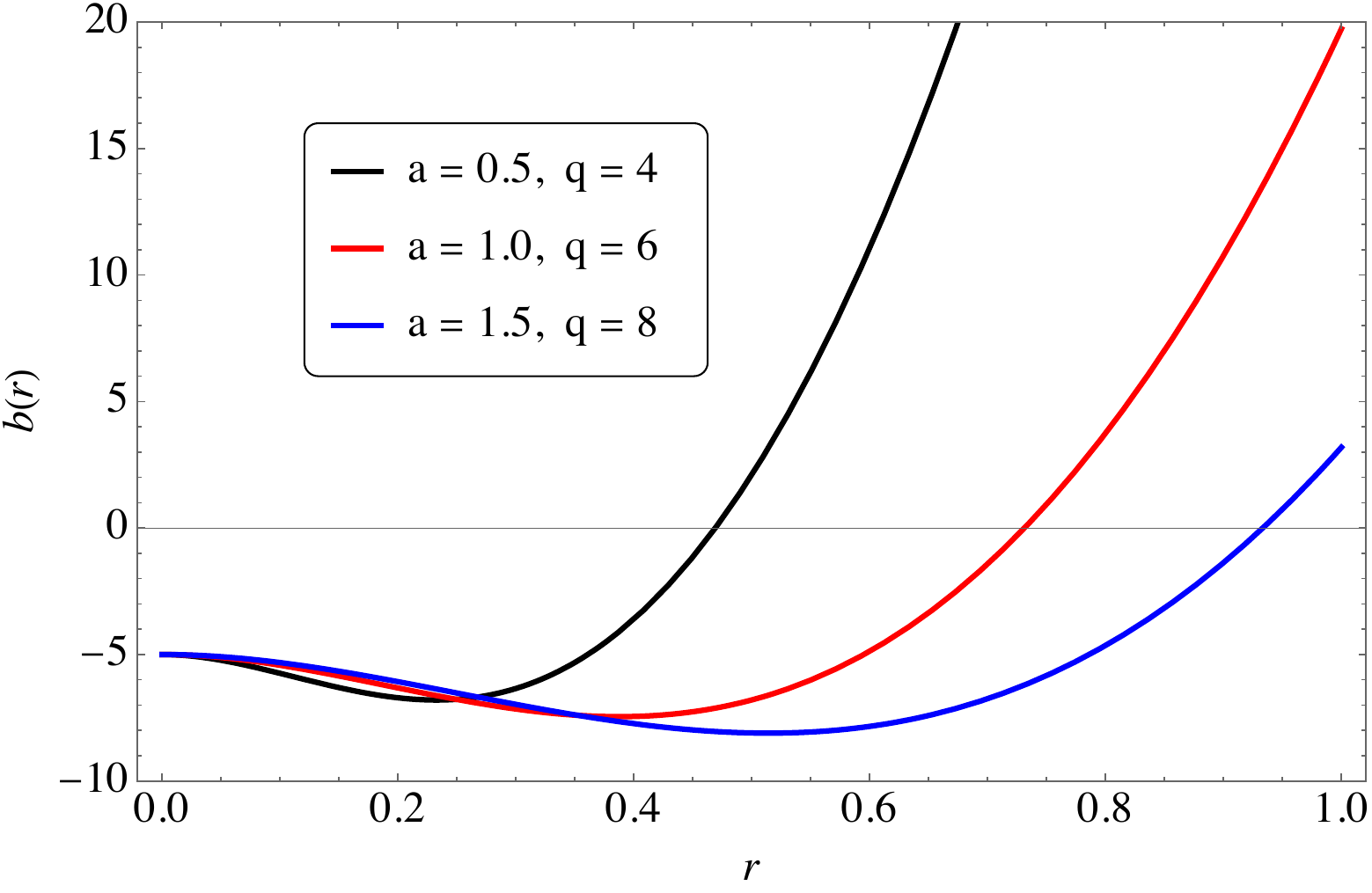}
\end{tabular}
\end{center}
\caption{Metric coefficient $b(r)$ from Eq. (\ref{bnew}), as a function of the radial coordinate $r$, for some values of $a$ and $q$, with $\Lambda=-20$, $M=5$ and $\kappa^2=1$ in natural units.}
\label{Figure1}
\end{figure}

For an event horizon to exist, the metric function $b(r)$ must have zeros. In the case of $\alpha=\beta=2$, the function $b(r)$ takes the values $-M$ at $r=0$ and diverges to $+\infty$ as $r\rightarrow+\infty$. Thus, it always has one and only one zero, which corresponds to the event horizon for this type of RBH. This situation is illustrated in Fig. \ref{Figure1} for different values of the parameters $a$ and $q$. 

Besides the particular cases considered above, exploring the limit $r\to \infty$ one can identify the general family of Lagrangians that smoothly tend to  Maxwell's electrodynamics in the far region. In fact, in this region, the Lagrangian (\ref{Lr}) tends to 
\begin{equation}
   \lim_{r\rightarrow+\infty}L(r)=\frac{2q^{2}}{a^{2}}\left(\frac{\Gamma\left(\frac{\beta+2}{\beta}\right)\Gamma\left(\frac{\alpha-1}{\beta}\right)}{\Gamma\left(\frac{\alpha+1}{\beta}\right)}-1\right).\label{Lrinf}
\end{equation}
It is now immediate to verify that for $\alpha=\beta+1$, the limit is exactly zero, whereas for $\alpha \neq \beta +1$, with $\alpha>1$ and $\beta>0$, the limit is a non-zero constant. 
Though the existence of a non-zero constant at infinity is somewhat arbitrary, as it can be absorbed in a redefinition of the cosmological constant, the point is that the $\alpha=\beta+1$ family of solutions admits important simplifications, leading to 
\begin{equation}\label{ELsol}
E(r)=\frac{qr^{\beta+1}}{\left(r^{\beta}+a^{\beta}\right)^{\frac{\beta+2}{\beta}}},\ \ \ \ L(r)=\frac{2q^{2}\left(r^{\beta}-a^{\beta}\right)}{\left(r^{\beta}+a^{\beta}\right)^{\frac{\beta+2}{\beta}}},
\end{equation}
and
\begin{equation}
b(r)=-M-\Lambda r^{2}-2\kappa^{2}q^{2}\frac{r^{2}}{a^{2}}\,_{2}F_{1}\left(\frac{2}{\beta},\frac{2}{\beta};\frac{\beta+2}{\beta};-\left(\frac{r}{a}\right)^{\beta}\right),\label{brnew2}
\end{equation}where $\beta>0$. 

Interestingly, the above solution can be employed to describe a broad spectrum of NED models found in the literature. In Table \ref{tab1} below, we present an incomplete list of models that can be recovered from our solutions (\ref{ELsol}) and (\ref{brnew2}). It is noteworthy to observe that for integer values of $\beta$ greater than 5 and for $\beta=3$, no solution exists in terms of elementary functions. On the other hand, when $\beta$ takes on rational values in the form $\frac{2}{n-1}$, with $n$ being a positive integer greater than 2, we identify a closed solution featuring the function $b(r)$ with a logarithmic term, such as $\ln \left[\left(\frac{r}{a}\right)^{\frac{2}{n-1}}+1\right]$. In all cases, these solutions represent regular black holes. Moreover, we explicitly verify that all solutions satisfy the energy conservation equation (\ref{Energy2}) when we set $\lambda=0$, which is just the case analyzed here.

\begin{table}[!h]  
\begin{center}
\resizebox{\textwidth}{!}{ 
\begin{tabular}{|c|c|c|c|c|} 
\hline
\hline  
$\beta$ & Electric field $E(r)$ & Lagrangian ${L(r)}$ & Metric function $b(r)$ & References \\ 
\hline
1 & $\frac{q r^2}{(r+a)^3}$ & $\frac{2 q^2 (r-a)}{(r+a)^3}$ & $-M-\Lambda r^{2}-4\kappa^{2}q^{2}\left[\ln\left(\frac{r}{a}+1\right)+\frac{a}{r+a}\right]$ & Case III \cite{He:2017ujy} \\
\hline
2 & $\frac{q r^3}{\left(r^2+a^2\right)^2}$ & $\frac{2 q^2 \left(r^2-a^2\right)}{\left(r^2+a^2\right)^2}$ & $-M-\Lambda r^2 -2 \kappa^2 q^2 \ln \left(\frac{r^2}{a^2}+1\right)$ & Case II \cite{Cataldo:2000ns,He:2017ujy} \\
\hline
4 & $\frac{q r^5}{\left(r^4+a^4\right)^{3/2}}$ & $\frac{2 q^2 \left(r^4-a^4\right)}{\left(r^4+a^4\right)^{3/2}}$ & $-M-\Lambda r^2 -2 \kappa^2 q^2 \text{arcsinh}\left(\frac{r^2}{a^2}\right)$ & Case IV \cite{He:2017ujy} \\
\hline
$\frac{1}{2}$ & $\frac{q r^{3/2}}{\left(\sqrt{r}+\sqrt{a}\right)^5}$ & $\frac{2 q^2 \left(\sqrt{r}-\sqrt{a}\right)}{\left(\sqrt{r}+\sqrt{a}\right)^5}$ & $-M-\Lambda r^2+\frac{4 \kappa^2 q^2 \sqrt{r} \left(15 \sqrt{a r}+6 a+11 r\right)}{3 \left(\sqrt{r}+\sqrt{a}\right)^3}-8 \kappa^2 q^2 \ln \left(\sqrt{\frac{r}{a}}+1\right)$ & Pinto et al.\\ 
\hline 
$\frac{2}{n-1}, \forall n \in \mathbb{N}, n>1$ & $\frac{q r^{\frac{n+1}{n-1}}}{\left(r^{\frac{2}{n-1}}+a^{\frac{2}{n-1}}\right)^{n}}$ & $\frac{2 q^2 \left(r^{\frac{2}{n-1}}-a^{\frac{2}{n-1}}\right)}{\left(a^{\frac{2}{n-1}}+r^{\frac{2}{n-1}}\right)^{n}}$ & Logarithmic term $\sim \ln \left[\left(\frac{r}{a}\right)^{\frac{2}{n-1}}+1\right]$ & Pinto et al.\\ 
\hline
\end{tabular}}
\caption{\label{tab1} Incomplete List of some NED models.}
\end{center}
\end{table}

\subsection{$f(R,T)$ solutions ($\lambda\neq 0$)}
Substituting the electric field (\ref{electricfield}) into Eq. (\ref{eq:L-Econstrain}) yields a second-order linear differential equation for the function $L(r)$. Unfortunately, the expression for $L(r)$ cannot be integrated exactly for arbitrary values of $\alpha$ and $\beta$, and its expression in terms of integrals is not very illuminating. Accordingly, and for concreteness, let us focus on the particular values for $\beta$ that we use in table \ref{tab1}. This facilitates the comparison with the GR case and allows us to see the effect of the $\lambda$-parameter on the Lagrangian density and on the $b(r)$ function.

For $\beta=1$, the electric field is given by
\begin{equation}\label{ELsol2}
E(r)=\frac{qr^{2}}{\left(r+a\right)^{3}},
\end{equation}
and one can use Eq. (\ref{eq:L-Econstrain}) to obtain the NED Lagrangian, which takes the form
\begin{align}
L(r) & =c_{1}+\frac{\left(2\kappa^{2}-\lambda\right)q^{2}(r-a)}{\left(\kappa^{2}+\lambda\right)(r+a)^{3}}-\frac{2c_{2}\lambda }{2\kappa^{2}+3\lambda}\left(\frac{r}{(r+a)^{\frac{3}{2}}}\right)^{3+\frac{2\kappa^{2}}{\lambda}}\nonumber \\
& +\frac{3\lambda q^{2}\left(2\kappa^{2}-\lambda\right)}{a\left(\kappa^{2}+\lambda\right)\left(2\kappa^{2}+\lambda\right)}\int^{r}\frac{r'(r'-2a)}{(r'+a)^{4}}\,_{2}F_{1}\left(1,\frac{\kappa^{2}}{\lambda}+\frac{1}{2};-\frac{2\kappa^{2}}{\lambda};-\frac{r'}{a}\right)dr'. \label{eq:Llambda}
\end{align}

 In the first line of this expression we have three terms with different properties. First, we find a constant $c_1$, which can be set to zero to recover Maxwell asymptotically. The second term recovers the GR limit when $\lambda\to 0$. The third term has been written in that way to make it clear that it vanishes when $\lambda\to 0$ because its dependence on $r$ is given by a positive number smaller than one which is raised to a power that diverges in the GR limit. Moreover, due to the convergence criterion of the hypergeometric function, we must have $\lambda < 0$, which prevents the pathological case where $\lambda = 2\kappa^{2} > 0$, as seen in Sec. \ref{review}.
 
 Interestingly, the constant $c_{2}$ does not have any physical implications because it does not appear in the differential equation for the function $b(r)$, whose form can be written as
\begin{align}
b(r) & =-M-\Lambda r^{2}-\frac{aq^{2}(2\kappa^{2}-\lambda)\left[2\kappa^{2}(r+a)+\lambda(7r+6a)\right]}{(\kappa^{2}+\lambda)(r+a)^{2}}-\frac{\left(4\kappa^{4}+8\kappa^{2}\lambda-5\lambda^{2}\right)}{\kappa^{2}+\lambda}q^{2}\ln\left(\frac{r+a}{a}\right)\nonumber \\
 & +\frac{3\lambda q^{2}(2\kappa^{2}-\lambda)}{a(\kappa^{2}+\lambda)(2\kappa^{2}+\lambda)}\int^{r}\frac{r''}{\left(r''+a\right)^{3}}\left[2\lambda r''^{2}\,_{2}F_{1}\left(1,\frac{\kappa^{2}}{\lambda}+\frac{1}{2};-\frac{2\kappa^{2}}{\lambda};-\frac{r''}{a}\right)\right.\nonumber \\
 & \left.-\left(2\kappa^{2}+3\lambda\right)\left(r''+a\right)^{3}\int^{r''}\frac{\left(2a-r'\right)r'}{\left(r'+a\right)^{4}}\,_{2}F_{1}\left(1,\frac{\kappa^{2}}{\lambda}+\frac{1}{2};-\frac{2\kappa^{2}}{\lambda};-\frac{r'}{a}\right)dr'\right]dr''.
\end{align}
Note that the first line in the above equation recovers the result in table \ref{tab1} when $\lambda\rightarrow0$. To guarantee a well-behaved numerical solution for $b(r)$, we must be able to numerically integrate the hypergeometric function $\,_{2}F_{1}\left(1,\frac{\kappa^{2}}{\lambda}+\frac{1}{2};-\frac{2\kappa^{2}}{\lambda};-\frac{r'}{a}\right)$. To do so, we must consider particular negative values of $\lambda$ that can turn the hypergeometric function into a simple polynomial function and not just arbitrary ones. Those values can be found through the equality between the second argument of the hypergeometric function and a positive integer number $n$. As such, in this case, those are given by $\lambda = - \frac{\kappa^2}{n + \frac{1}{2}}$.
\begin{figure}[!h]
\begin{center}
\begin{tabular}{ccc}
\includegraphics[height=7cm]{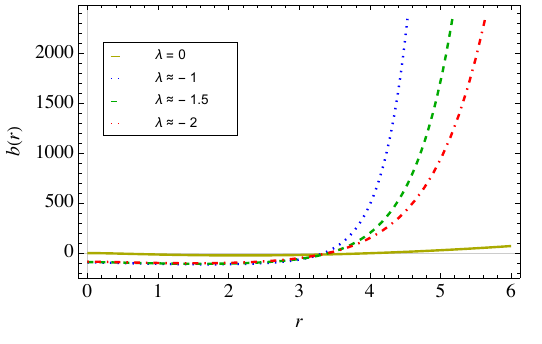}
\end{tabular}
\end{center}
\caption{Metric coefficient function $b(r)$ as a function of the radial coordinate, for three different values of $\lambda$, for $\beta=1$. Here, we fixed $a=1$, $q=1$, $\kappa^2=8 \pi$, $M=1$ and $\Lambda = - 5$ in natural units.}
\label{metricbeta1}
\end{figure}

\begin{figure}[!ht!]
\begin{center}
\begin{tabular}{ccc}
\includegraphics[height=5.2cm]{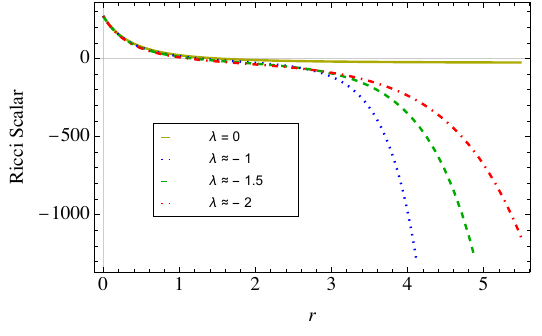} 
\includegraphics[height=5.2cm]{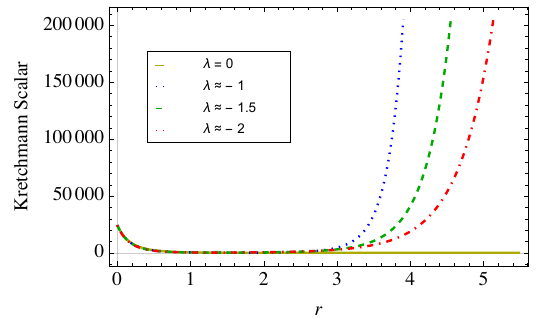}\\
(a)\hspace{4.7cm}(b)\\
\end{tabular}
\end{center}
\vspace{-0.5cm}
\caption{Ricci scalar (left panel) and Kretschmann scalar (right panel) as functions of the radial coordinate, for four different values of $\lambda$, in the case $\beta = 1$. For both plots, we fixed $a=1$, $q=1$, $\kappa^2=8 \pi$ and $\Lambda = - 5$ in natural units.}
\label{curvature scalars 1}
\end{figure}

\begin{figure}[!h]
\begin{center}
\begin{tabular}{ccc}
\includegraphics[height=7cm]{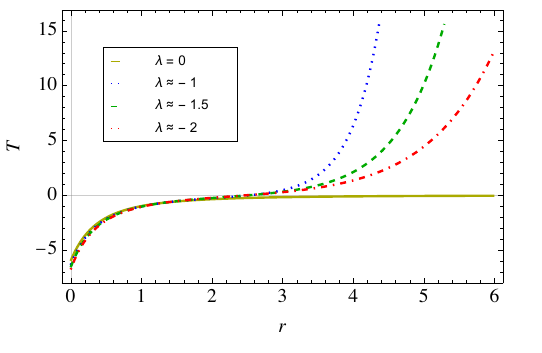}
\end{tabular}
\end{center}
\caption{Trace of the energy-momentum tensor $T$ as a function of the radial coordinate, for three different values of $\lambda$, for $\beta=1$. Here, we fixed $a=1$, $q=1$, and $\kappa^2=8 \pi$ in natural units. This quantity is independent of $M$ and $\Lambda$.
\label{tracebeta1}}
\end{figure}

In Fig. \ref{metricbeta1}, we plotted the resulting metric coefficient function $b(r)$ after numerical integration. For all considered values of $\lambda$, we found that this RBH solution possesses one event horizon, and its position varies with the value of $\lambda$ in a non-trivial manner for small values of $r$. Nonetheless, for $\lambda\neq0$, it is clear that for smaller values of $\lambda$ the function $b(r)$ grows more rapidly. Furthermore, in Fig. \ref{curvature scalars 1} we computed the Ricci and Kretschmann scalars as functions of the radial coordinate $r$. On the one hand, we verify that the curvature scalars remain regular (finite) throughout the radial coordinate range for $\lambda=0$. On the other hand, for all values $\lambda \neq 0$, both scalars diverge when $r \rightarrow \infty$. Curiously, this means that the further we are from the origin of the RBH, the effects of gravity become more intense.  In order to better understand this behavior, in Fig. \ref{tracebeta1} we plotted the trace of the energy-momentum tensor as a function of $r$, which shows that $T$ is a monotonic function that behaves similarly as the curvature scalars: in the case of GR ($\lambda = 0$) it reaches a plateau while for $f(R,T)$ gravity ($\lambda \neq 0 $), it diverges when $r \rightarrow \infty$. In fact, a fast growth of $T$ occurs near the horizon, meaning that outside of the RBH these contributions play an important role. Indeed, there is a clear correlation between the divergent growth of $T$ and that of the curvature scalars.

\begin{figure}[!h]
\begin{center}
\begin{tabular}{ccc}
\includegraphics[height=7cm]{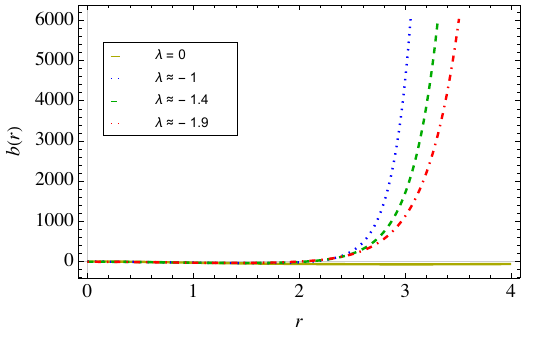}
\end{tabular}
\end{center}
\caption{Metric coefficient function $b(r)$ as a function of the radial coordinate, for three different values of $\lambda$, for $\beta=2$. Here, we fixed $a=1$, $q=1$, $\kappa^2=8 \pi$, $M=1$ and $\Lambda = - 5$ in natural units.
\label{metricbeta2}}
\end{figure}

\begin{figure}[!ht!]
\begin{center}
\begin{tabular}{ccc}
\includegraphics[height=5.2cm]{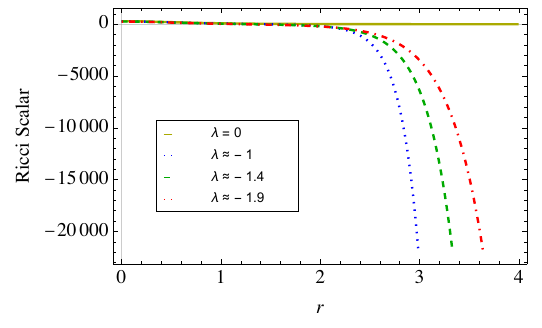} 
\includegraphics[height=5.2cm]{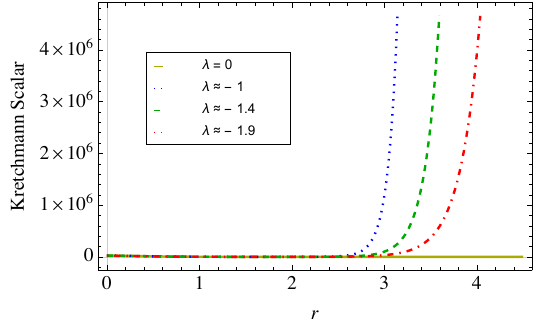}\\
(a)\hspace{4.7cm}(b)\\
\end{tabular}
\end{center}
\vspace{-0.5cm}
\caption{Ricci scalar (left panel) and Kretschmann scalar (right panel) as functions of the radial coordinate, for four different values of $\lambda$, in the case $\beta = 2$. For both plots, we fixed $a=1$, $q=1$, $\kappa^2=8 \pi$ and $\Lambda = - 5$ in natural units.}
\label{curvature scalars 2}
\end{figure}

\begin{figure}[!h]
\begin{center}
\begin{tabular}{ccc}
\includegraphics[height=7cm]{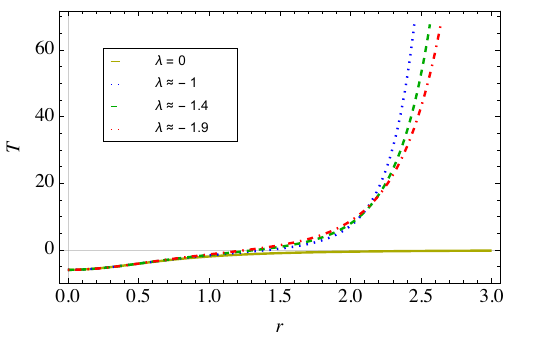}
\end{tabular}
\end{center}
\caption{Trace of the energy-momentum tensor $T$ as a function of the radial coordinate, for three different values of $\lambda$, for $\beta=2$. Here, we fixed $a=1$, $q=1$, and $\kappa^2=8 \pi$ in natural units. This quantity is independent of $M$ and $\Lambda$.
\label{tracebeta2}}
\end{figure}

For $\beta=2$, the electric field is given by
\begin{equation}\label{ELsol3}
E(r)=\frac{qr^{3}}{\left(r^2+a^2\right)^{2}},
\end{equation}
and Eq. (\ref{eq:L-Econstrain}) leads to

\begin{align}
L(r) & =c_{1}+\frac{2q^{2}(r^{2}-a^{2})}{\left(r^{2}+a^{2}\right)^{2}}+\frac{2c_{2}\lambda }{2\kappa^{2}+3\lambda}\left(\frac{r^{\frac{3}{2}}}{\left(r^{2}+a^{2}\right)}\right)^{3+\frac{2\kappa^{2}}{\lambda}}\nonumber \\
 & +\frac{32\lambda q^{2}}{6\kappa^{2}+5\lambda}\int\frac{r\left(r^{2}-3a^{2}\right)}{\left(r^{2}+a^{2}\right)^{3}}\,_{2}F_{1}\left(-\frac{3\kappa^{2}}{2\lambda}-\frac{5}{4},-\frac{2\kappa^{2}}{\lambda};-\frac{(6\kappa^{2}+\lambda)}{4\lambda};-\frac{r^{2}}{a^{2}}\right)\,dr \ .
\end{align}
For the function $b(r)$, the solution can be written as
\begin{align}
b(r) & =-M-\Lambda r^{2}+\frac{4\lambda q^{2}a^{2}}{r^{2}+a^{2}}-\left(2\kappa^{2}-\lambda\right)q^{2}\ln\left(\frac{r^{2}}{a^{2}}+1\right)\nonumber \\
 & +\int^{r}\left[\frac{64\lambda^{2}q^{2}r''^{3}}{\left(6\kappa^{2}+5\lambda\right)\left(r''^{2}+a^{2}\right)^{2}}\,_{2}F_{1}\left(1,\frac{\kappa^{2}}{2\lambda}-\frac{1}{4};-\frac{(6\kappa^{2}+\lambda)}{4\lambda};-\frac{r''^{2}}{a^{2}}\right)\right.\nonumber \\
 & \left.+\left(2\kappa^{2}+3\lambda\right)r''\int^{r''}\frac{32\lambda q^{2}r'\left(r'^{2}-3a^{2}\right)}{\left(6\kappa^{2}+5\lambda\right)\left(r'^{2}+a^{2}\right)^{3}}\,_{2}F_{1}\left(1,\frac{\kappa^{2}}{2\lambda}-\frac{1}{4};-\frac{(6\kappa^{2}+\lambda)}{4\lambda};-\frac{r'^{2}}{a^{2}}\right)dr'\right]dr''.
\end{align}
As expected, the above equations recover the result in table \ref{tab1} when $\lambda\rightarrow0$. Besides, to guarantee the convergence of the hypergeometric function, we must have $-\frac{6 \kappa^{2} + \lambda}{4\lambda} > 0$, which implies that $-6 < \lambda/\kappa^{2} < 0$. Thus, we observe an additional restriction on the interval of the $\lambda$-parameter, and it remains negative. Additionally, as we did in the $\beta = 1$ case, we consider particular values of $\lambda$ that make the integration of the hypergeometric function smooth. This time, those are given by $\lambda = - \frac{\kappa^2}{2n -\frac{1}{2}}$, with $-6\kappa^{2} < \lambda < 0$. The resultant $b(r)$ function is plotted in Fig. \ref{metricbeta2}. As in the case of $\beta = 1$, the RBH solution for $\beta = 2$ is characterized by only one event horizon. Additionally, as the absolute value of $\lambda$ increases, the event horizon is located closer to the origin, with the case of $\lambda=0$ being the most extreme, as it only has it at $r \approx 6.0568$. However, for $r\gtrsim 2.45$, $b(r)$ becomes bigger more rapidly for smaller values of $\lambda$, except for $\lambda=0$. Moreover, the corresponding Ricci and Kretschmann scalars are plotted in Fig. \ref{curvature scalars 2}. As before, the main consequence of introducing the $\lambda T$ term is the divergence of both scalars in the limit $r \rightarrow \infty$. Figure \ref{tracebeta2} further sustains this conclusion, as we verify that the trace of the energy-momentum tensor also goes to infinity when $\lambda\neq0$ in the case of $\beta=2$.

\subsubsection{Non-conservation of energy-momentum}

\begin{figure}[!ht!]
\begin{center}
\begin{tabular}{ccc}
\includegraphics[height=5.5cm]{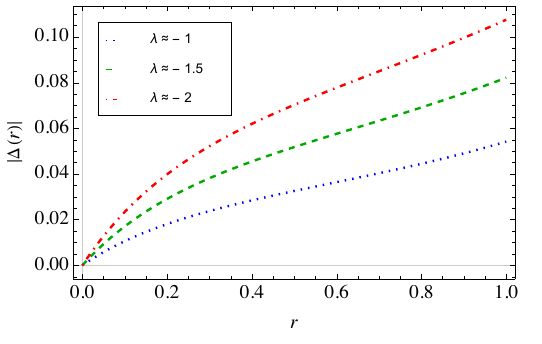} 
\includegraphics[height=5.5cm]{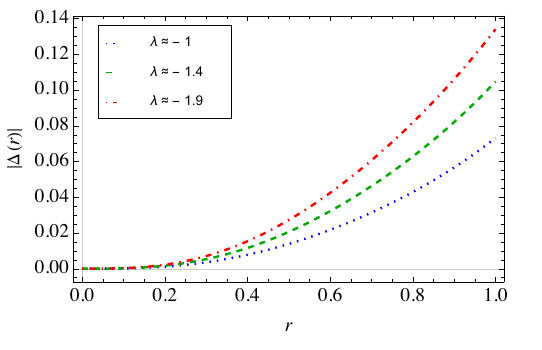}\\
(a)\hspace{6.7cm}(b)\\
\end{tabular}
\end{center}
\vspace{-0.5cm}
\caption{Deviation from pure energy-momentum conservation equation evaluated from Eq. (\ref{desviation}), as a function of the radial coordinate, for three different values of $\lambda$, and for $\beta=1$ (left panel) and $\beta=2$ (right panel). For both plots, we fixed $a=1$, $q=1$, and $\kappa^2=8 \pi$ in natural units.}
\label{deviation}
\end{figure}

Finally, we discuss the (non) energy-momentum conservation for $f(R,T)$ gravity in the present context. To evaluate the deviation from the energy-momentum conservation equation (\ref{Energy2}), we analyze the quantity  $\Delta$, defined as  
\begin{equation}
   \Delta \equiv \frac{\lambda}{\kappa^2-\lambda}\nabla^{\mu}\left(\Theta_{\mu\nu}-\frac{1}{2}g_{\mu\nu}T\right),\label{desviation}
\end{equation}  
where, of course, only the nonzero component is taken into account. This quantity is plotted as a function of $r$, for $\beta=1$ and $\beta=2$, in Fig. \ref{deviation}.

By analyzing Fig. \ref{deviation}, we verify that the absolute value of $\Delta$ is a monotonic function of the radial coordinate $r$  for all considered values of $\lambda$ in  both $\beta=1$ and $\beta=2$ cases. Furthermore, the bigger the absolute value of $\lambda$, the bigger the deviation from energy-momentum conservation, which is in accordance with Eq. \eqref{desviation}. In addition, while for $\beta = 1$ the deviation has a more accentuated growth near the origin and then grows linearly for bigger values of $r$, for $\beta = 2$ the deviation is more accentuated far from the origin, being almost 0 until $r \approx 0.2$ for all values of $\lambda$. From the physical point of view, both cases demonstrate signs of cumulative non-conservative energy-momentum processes along the radial coordinate that are induced by the $\lambda T$ term. Such processes may be photon self-interactions, which, although impossible in the context of the Standard Model of Particle Physics, are theoretically possible in the framework of NEDs. These processes may be the cause of why the curvature scalars seem to diverge in the limit $r\rightarrow \infty$ when $\lambda \neq 0$, in both $\beta=1$ and $\beta=2$ cases.

\section{Summary and conclusion \label{conclusion}}
In this paper, we investigated the existence of regular black hole solutions in (2+1)-dimensional $f(R, T)$ gravity coupled to nonlinear electrodynamics. We started by assuming that the form of the $f(R, T)$ function consists of GR + $\Lambda$  plus a linear term on the trace of the energy-momentum tensor, $\lambda T$, so that the modified gravitational dynamics could be kept to a minimum. Additionally, we assumed a static and circularly symmetric space-time dependent on a function $b(r)$, and we neglected the possible magnetic field contributions to the Faraday-Maxwell tensor. With these assumptions in mind, and demanding that one recovers the usual form for the electric field $E(r)=q/r$ if the Lagrangian is $L(F)=-F$, we obtained the modified gravitational field equations and the modified gauge field equations. Since we only had two equations for three variables, $E(r)$, $L(r)$ and $b(r)$, we opted to propose a generalized Maxwell-like form for $E(r)$, with a dependence on a characteristic distance, $a$, and two parameters, $\alpha$ and $\beta$. We verified that for $\alpha=a=0$ and $\beta=1$ (the exact Maxwell electric field, $E(r)=q/r$), we could obtain the static charged and static BTZ black hole solution \cite{Banados:1992wn,Banados:1992gq}.

In order to obtain the general solutions for the Lagrangian $L(r)$ and metric function $b(r)$ related to the electric field $E(r)$ we proposed, we first considered the GR case ($\lambda = 0$) for simplicity. By attributing specific values to the parameters $\alpha$ and $\beta$, not only did we found new solutions but also rediscovered older ones. Attending to the behavior of the electromagnetic Lagrangian in the far asymptotic region, we identified a family of NEDs for which Maxwell's electrodynamics is smoothly recovered. This allowed us to write the electric field, the Lagrangian, and the metric function in terms of just two parameters instead of three. We confirmed that this set of solutions constitutes a family of regular black holes in (2+1)-dimensional GR that satisfy the Maxwell behavior in the asymptotic far limit, which encompasses several particular cases previously found in the literature \cite{Cataldo:2000ns, He:2017ujy} ($\beta=1,2,4$).

Furthermore, we provided the first regular black hole solutions in (2+1)-dimensional $f(R, T)$ gravity ($\lambda \neq 0$). Unfortunately, the  non-linearity of the equations prevented us from obtaining a whole class of general solutions as obtained in the GR limit. Thus, to circumvent this issue, we constructed particular solutions by considering specific values for the $\beta$ parameter. In fact, for $\beta=1$ and $\beta = 2$, we found that it is possible to obtain an expression for both the Lagrangian $L(r)$ and the metric coefficient function $b(r)$ that is the GR part plus a contribution due to $T$. In addition, we computed the corresponding Ricci and Kretschmann scalars, as well as $T$, and we evaluated the non-conservation of energy and momentum, all in terms of the radial distance $r$. For $\lambda = 0$, we have finite curvature scalars and conservation of energy and momentum, while for $\lambda \neq 0$, such scalars diverge in the limit $r \rightarrow \infty$. The origin of this divergence is difficult to track in general from the formulas obtained due to their complexity, though it is most likely found in the hypergeometric functions, which may introduce powers of $r$ that make the resulting $L(r)$ grow with the radius. In such cases, since $T$ and $\Theta_{\mu\nu}$ are directly affected by the form of $L(r)$ and its derivatives, residual terms that grow with $r$ may induce the observed growths, regardless of the fact that the electric field is designed to be well-behaved and decaying everywhere. This justifies the correlated growth in the scalars and the non-conservation of energy-momentum away from the black hole. Though exceptional cases with smooth asymptotics may exist, we can conclude that the general trend is to intensify the strength of gravity at far distances despite describing regular black holes.

To conclude, the RBH geometries described here in the framework of (2+1)-dimensional $f(R, T)$ gravity indicate that pure analytical solutions of this nature in the (3+1)-dimensional version of the same theory may also exist and are worthy of investigation. Whether such solutions also develop a generic increase in the strength of gravity in the far region is an aspect that will be explored in the future.


-------------------------------------------------------------------------

\section*{Acknowledgments}
\hspace{0.5cm} RVM thank the Coordena\c{c}\~{a}o de Aperfei\c{c}oamento de Pessoal de N\'{i}vel Superior (CAPES), and the Conselho Nacional de Desenvolvimento Cient\'{i}fico e Tecnol\'{o}gico (CNPq), Grants no
311732/2021-6 (RVM) and 308268/2021-6 (CRM), for financial support. MASP acknowledge support from the FCT - Funda\c{c}\~{a}o para a Ci\^{e}ncia e a Tecnologia, I.P. research grants UIDB/04434/2020 and UIDP/04434/2020, and through the FCT project with reference PTDC/FIS-AST/0054/2021 (``BEYond LAmbda''). MASP also acknowledges support from the FCT through the Fellowship UI/BD/154479/2022 (https://doi.org/10.54499/UI/BD154479/2022). The authors also acknowledge financial support from the Spanish Grants  PID2020-116567GB-C21, PID2023-149560NB-C21 funded by MCIN/AEI
/10.13039/501100011033, and by CEX2023-001292-S funded by MCIU/AEI.  The paper is based upon work from COST Actions CosmoVerse CA21136 and CaLISTA CA21109 supported by COST (European Cooperation in Science and Technology).


\end{document}